\documentclass[preprint,review,12pt]{elsarticle}

\usepackage{amssymb}

\newcommand{\emeas}{$E_\textrm{\scriptsize meas}$}
\newcommand{\erecoil}{$E_\textrm{\scriptsize recoil}$}

\begin{document}

\begin{frontmatter}

\title{Measurement of the quenching and channeling effects\\ in a CsI crystal used for a WIMP search}

\author[label1,label9]{J. H. Lee\corref{cor1}}
\cortext[cor1]{juheelee1978@gmail.com}
\author[label1]{G. B. Kim}
\author[label1]{I. S. Seong}
\author[label1]{B. H. Kim}
\author[label1]{J. H. Kim}
\author[label1]{J. Li}
\author[label1]{J. W. Park}
\author[label1]{J. K. Lee}
\author[label1]{K. W. Kim}
\author[label1]{H. Bhang}
\author[label1]{S. C. Kim}
\author[label1]{Seonho Choi\corref{cor2}}
\cortext[cor2]{choi@phya.snu.ac.kr}
\author[label1]{J. H. Choi}
\author[label1]{H. W. Joo}
\author[label1]{S. J. Lee}
\author[label1]{S. L. Olsen}
\author[label1]{S. S. Myung}
\author[label1]{S. K. Kim} 
\author[label8,label2]{Y. D. Kim}
\author[label2]{W. G. Kang}
\author[label3]{J. H. So}
\author[label3]{H. J. Kim}
\author[label4]{H. S. Lee}
\author[label4]{I. S. Hahn}
\author[label5]{D. S. Leonard}
\author[label6]{J. Li}
\author[label6]{Y. J. Li}
\author[label6]{Q. Yue}
\author[label7]{X. R. Li}
\address[label1]{Seoul National University, Seoul 151 - 747, Korea}
\address[label9]{Korea Research Institute of Standards and Science, Daejeon 305-340, Korea}
\address[label8]{Center for Underground Physics, Institute for Basic Science, Daejeon 305-811, Korea}
\address[label2]{Sejong University, Seoul 143 - 747, Korea}
\address[label3]{Kyungpook National University, Daegu 702 - 701, Korea}
\address[label4]{Ewha womans University, Seoul 120 - 750, Korea}
\address[label5]{University of Seoul, Seoul 130 - 743, Korea}
\address[label6]{Tsinghua University, Beijing 10084, China}
\address[label7]{Institute of High Energy Physics (IHEP), Beijing 100049, China}

\begin{abstract}
We have studied channeling effects in a Cesium Iodide (CsI) crystal that is similar in composition to the ones being used in a search for Weakly Interacting Massive Particles (WIMPs) dark matter candidates, and measured its energy-dependent quenching factor, the relative scintillation yield for electron and nuclear recoils. The experimental results are 
reproduced with a GEANT4
simulation that includes a model of the scintillation efficiency as a function of electronic stopping power. We present the measured and simulated quenching factors and the estimated effects of channeling.
\end{abstract}


\end{frontmatter}

\section{Introduction}
\label{sec:intro}

Weakly Interacting Massive Particles (WIMPs) are candidates for dark matter~\cite{wimp} that satisfy
the relic density of dark matter observed by the WMAP~\cite{WMAP} and PLANCK~\cite{PLANCK} experiments and
successfully explain the observed gravitational lensing effects of galatic clusters~\cite{jee, bradac}.
In the Standard Dark Matter Halo Model, WIMPs are spread throughout the galaxy with a Maxwellian velocity distribution
and transfer their kinetic energy to ordinary matter by WIMP-nucleon elastic scattering~\cite{SHM,lewin}.
The scattered nuclei subsequently produce detectable responses in the material via ionization, scintillation
and phonon creation, from which WIMP-nucleon scattering can be inferred.

Elastic scattering of WIMPs with masses of hundreds or thousands of GeV on nuclei would produce recoil nuclei with
energies ranging from tens to hundreds of keV.  Since it is hypothesized to be a weak interaction process, it is expected to occur with a very small probability. Thus, a WIMP detector is required to have a high
sensitivity for nuclear recoils and a very low level of radioactive background contaminants.
Thallium-doped CsI crystals (CsI(Tl)) are well suited for WIMP searches thanks to their high scintilation light
yields, about 65,000 photons/MeV for electron recoils~\cite{skkim, hjkim}, and the low radioactive background levels
that can be achieved through purification techniques applied during the growing process~\cite{ydkim, hslee_NIMA}.
A particular advantage of CsI(Tl) detectors is their different time responses for electron- and nuclear-induced
ionization, which permits the use of pulse shape discrimination (PSD) techniques to remove significant fractions of electron-recoil
background events induced by radioactive background sources~\cite{hspark, hslee_JINST}.  Spin-independent WIMP-nucleus elastic scattering
is a coherent process with a cross section that has quadratic dependence on the atomic mass number; the cross section for spin-dependent elastic scattering
has a quadratic dependence on the nuclear spin expectation value~\cite{lewin}.  Because of their large atomic mass
numbers and nuclear-spin expectation values, Cesium (Cs) and Iodine (I) ions are expected to have relatively large WIMP-nucleus
cross sections for both the spin-independent and spin-dependent scenarios. The Korea Invisible Mass Search (KIMS) uses CsI(Tl) crystals
for WIMP searches and has published stringent upper limits for WIMP-proton spin-dependent
elastic scattering cross-sections~\cite{hslee_PRL, sckim_PRL}.

The quenching factor (QF) is the scintillation light yield produced by a nuclear recoil relative to that for an electron recoil
at the same energy.  Typically electron-recoil responses are measured using gamma-ray sources, in which case the QF can be expressed as:
\begin{equation}
 \textrm{QF} = \frac{E_\textrm{\scriptsize meas}}{E_\textrm{\scriptsize recoil}} = \frac{L_\textrm{\scriptsize meas}}
{E_\textrm{\scriptsize recoil}}\frac{E_{\gamma,\textrm{\scriptsize calib}}}{L_{\gamma,\textrm{\scriptsize calib}}},
\label{eq:QF}
\end{equation}
where \erecoil{}, \emeas{}  are the recoil energy of the nucleus produced by WIMP-nucleus scattering and the experimentally
measured energy by the crystal detector, respectively. Since the light output, $L_\textrm{\scriptsize meas}$, is measured in the scintillator, $E_\textrm{\scriptsize meas}$ can be obtained from the calibration. $L_{\gamma, \textrm{\scriptsize calib}}$ is the measured scintillation light
yield for gamma rays with a known energy $E_{\gamma, \textrm{\scriptsize calib}}$.

The quenching factors of CsI crystals used in the KIMS experiment have been previously measured~\cite{hspark}. Subsequently,
it has been suggested that channeling effects in the scintillation crystal might enhance the quenching factors 
to as high as QF$\simeq 1$ for some specific nuclear-recoil conditions~\cite{bernabei_channeling}.  Channeling occurs when a recoil ion in the target material moves in a direction that is within a critical angle from a symmetry axis
or plane in the crystal lattice~\cite{bernabei_channeling, gemmel}.  In these cases, the recoil ion primarily loses
energy via numerous scatterings with atomic electrons around target nuclei that are confined to small scattering angles
because of the relatively large impact parameters between the moving ion and target nuclei. As a result, channeling effects show up as enhanced ion penetration ranges and larger numbers of electron-hole pairs in the target material with a resultant smaller stopping power. The large penetration range enhances the scintillation yield in the crystal in accordance with Birks' formula~\cite{birks,hitachi,tretyak}.

   However, the aforementioned channeling effects have only been
studied using ions that are incident from outside of the crystal into the empty space between symmetrically aligned lattice
atoms~\cite{gemmel}.  Thus, conclusions from previous studies of channeling effects might not apply to recoil ions
from WIMP-nucleus scattering, which occur inside the crystal.  In these cases, recoil ions, originally located at a crystal lattice point, initially travel at
a large angle from the adjacent target nuclei located near the symmetry axis or crystal plane because of a small initial impact parameter. As a result, the recoil ions cannot easily channel
through the empty space near the symmetric axis; this is known as the blocking effect.  Bozorgnia et al.~\cite{bozorgnia, gondolo}
point out that recoiling lattice ions have some chance to be channeled through the symmetric axis due to thermal lattice vibrations.
Nevertheless, the fraction of full channeling for isotropically scattered recoil ions is expected to be below 10~\% for CsI crystals~\cite{gondolo}.

In this report, we present new measurements of quenching factors for a monocrystalline CsI(Tl) crystal.  In addition, we provide estimates
of channeling effects in the crystal by comparing the measured energy spectrum with the result of the GEANT4 simulations coupled with a program called MARLOWE~\cite{marlowe,robinson}. The impinging
neutrons in the experimental setup were modelled with a GEANT4-based simulation and the propagation of the recoil ions in the monocrystalline
structure was incorporated using the MARLOWE program.

\section{Experiment}
\label{sec:exp}

\begin{figure}[!ht]
\begin{center}
\includegraphics[width=1.0\textwidth,keepaspectratio=true]{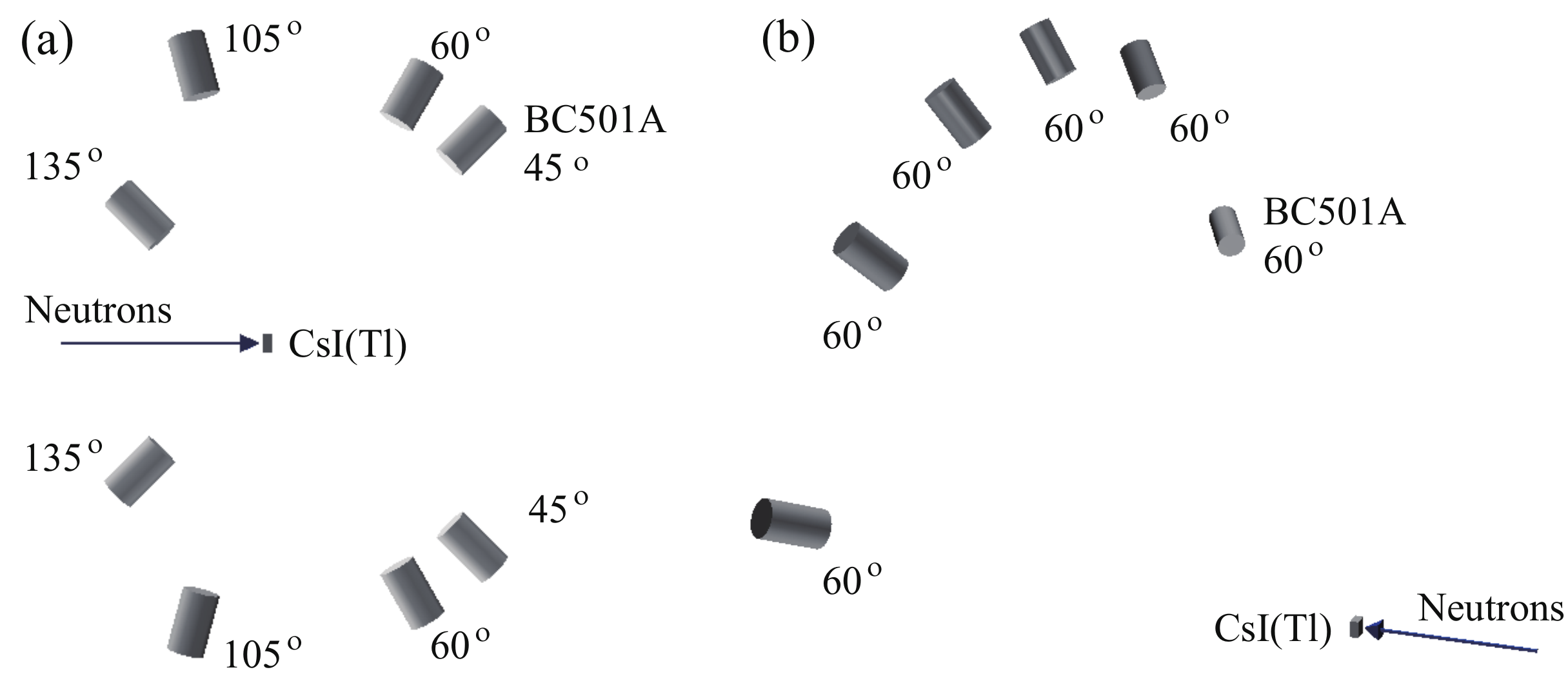}
\caption[] {Schematics of the experimental setups. (a) Setup I and (b) Setup II.
The arrows indicate the direction of the incident neutron beam. The numbers next to the neutron detectors represent polar scattering angles, $\Theta_\textrm{\scriptsize ND}$.
The shielding and supporting materials surrounding detectors are not drawn for clarity.}
\label{fig_setup}
\end{center}
\end{figure}

To study nuclear recoil signals in CsI(Tl), mono-energetic 2.4~MeV neutrons from a neutron generator were
irradiated onto a CsI crystal. A model MP320 neutron generator from ThermoFisher Scientific produces
neutrons isotropically from $^2$H($^2$H,n)$^3$He reactions. The maximum neutron flux is $7\times10^{5}$~/s for a 60~$\mu$A
beam current of 90~keV deuterons.  However, for these measurements we used 50~keV deuterons, for which the detected flux
of neutrons in a test experiment that placed a neutron detector at the output of the neutron generator was maximum, while the gamma flux was minimized. 

In order to measure the quenching factor and study channeling effects, two different experimental setups were used.  
The first one, called Setup I and depicted in Fig.~\ref{fig_setup}-(a), was optimized for measuring quenching factors in
the crystal for various neutron scattering angles, $\Theta_\textrm{\scriptsize ND}$, where the subscript ND indicates the neutron detector. In this setup, neutron detectors were located at eight different
polar scattering angles, 45$^{\circ}$, 60$^{\circ}$, 75$^{\circ}$, 90$^{\circ}$, 105$^{\circ}$, 120$^{\circ}$, 135$^{\circ}$, and 150$^{\circ}$. To increase
statistics, at each scattering angle, two neutron detectors were placed symmetrically on both sides of the neutron beam direction. For
90$^{\circ}$, to avoid blocking of the scattered neutrons by the PMT attached to the CsI crystal, the neutron detectors were
positioned in the plane perpendicular to the direction of the neutron. 
The distances from the CsI crystal to the neutron detectors ranged from 15.5~cm to 40.0~cm in order to partially compensate for the variation in cross sections of the neutron elastic scattering at the different scattering
angles.

The second setup, Setup II, described in detail in ref.~\cite{jhlee} and depicted in Fig.~\ref{fig_setup}-(b),
is designed to study channeling effects in nuclear recoil events. This has six neutron detectors at a 60$^{\circ}$ polar angle that cover different azimuthal recoil-ion directions. 
The six neutron detectors are  located 1.1~m away from the CsI crystal in order to restrict \erecoil{} values to a narrow range.

The CsI crystal had PMTs (9269QA by Electron Tubes, green-extended) attached at each end and was contained in a 1~mm thick
copper box.  The neutron detectors consisted of BC501A liquid scintillator in a 5~mm thick double-layered-bottle of teflon
and stainless steel read out by a R329-02 PMT made by HAMAMATSU. The CsI crystal dimensions were
$3\times3\times1.4$~cm$^{3}$ and the Thallium doping levels were under 0.1~\% mole.

The neutron generator was surrounded by a 10~cm thick polyethylene shield with an additional 10~cm thick lead wall on the side
facing the CsI crystal. Neutrons passing through a 15~cm thick-lead guide from the generator exited the generator via 3.2~cm
diameter apertures in the polyethylene and lead walls. In Setup I, to reduce ambient gamma backgrounds, 5~cm thick lead blocks were put around the
neutron detectors and the CsI crystal except along the neutron trajectories. The neutron detectors in Setup II were contained in 10~cm-thick polyethylene boxes to shield them from ambient radioactivity. In this setup, the CsI detector was not shielded along the neutron entrance and exit directions.

Events were triggered by a coincidence between valid signals from the CsI crystal and any one of neutron detectors in Setup I that occurs 
within a 2~$\mu$s timing window. For Setup II, an additional signal from the neutron generator was also required in order to reduce random coincident events.  A valid signal in the CsI crystal was defined as at least one photoelectron in each PMT
within a 2~$\mu$s interval.   This was realized with a 400~MHz Fast Analog to Digital Converter (FADC400) module made by Notice
Korea. The thresholds of the neutron detectors were set at the single photoelectron level.

\begin{figure}[!htb]
\begin{center}
\includegraphics[width=0.9\textwidth,keepaspectratio=true]{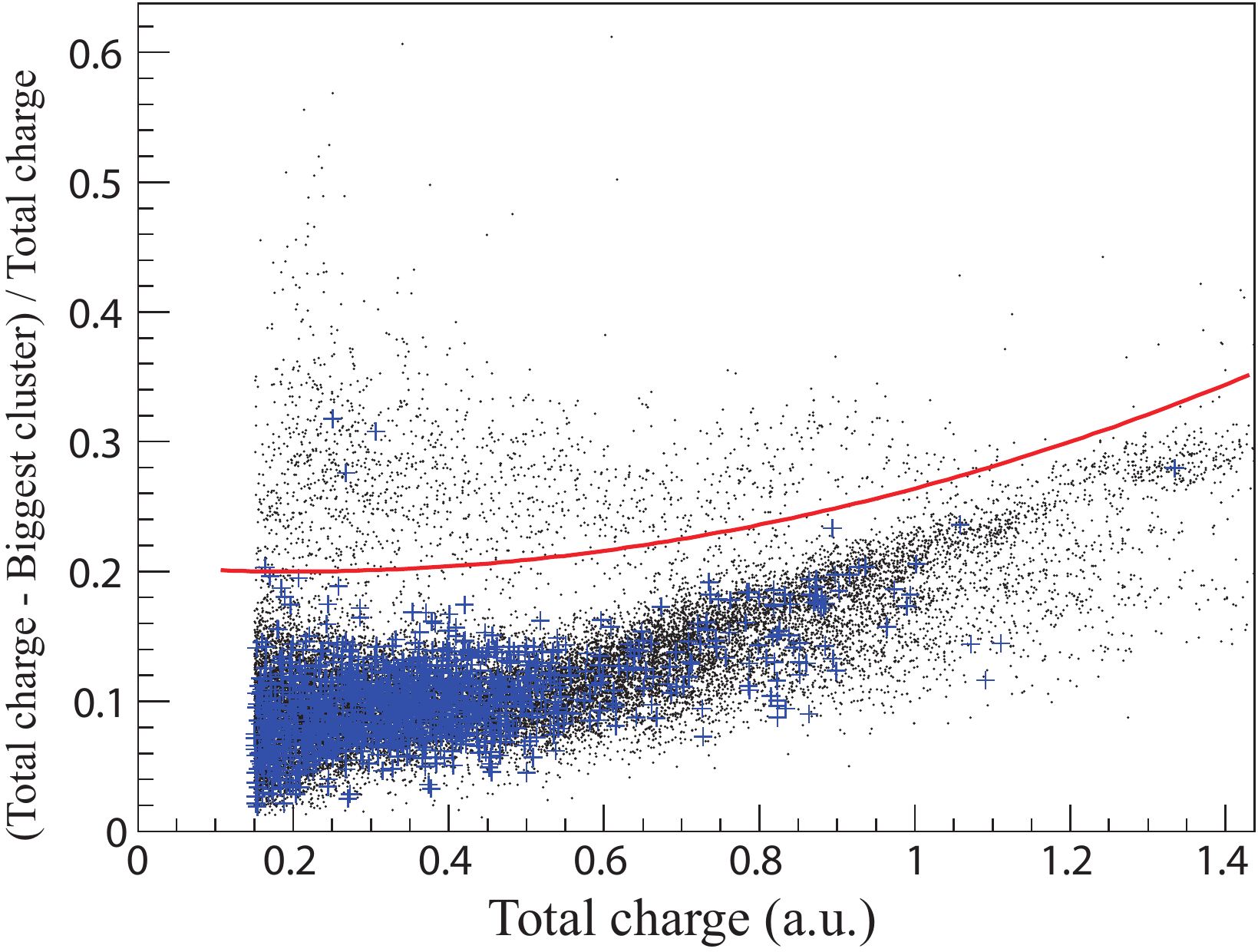}
\caption[] {(color online) A two-dimensional plot showing the pulse shape discrimination in one of the neutron detectors. The $x$-axis shows the
sum of all the collected charges for the first 1~$\mu$s time window. The $y$-axis shows the fraction of the charges not contained 
the cluster with the largest summed charge. The black dot points are from the neutron scattering experiment and the blue crosses are
from gamma ray calibration data using a $^{22}$Na source. Points above the red solid curve are identified as neutron-induced events; points below
the curve are rejected as gamma-induced events. }

\label{fig_neutron}
\end{center}
\end{figure} 

In the offline analysis, neutron scattering events were selected by exploiting the pulse shape discrimination (PSD) power of the
liquid scintillator as shown in the scatter plot of Fig.~\ref{fig_neutron}, where the horizontal axis shows the summed charge
over 1~$\mu$s and the vertical axis shows the fraction of the charge that is not in the dominant peak, which we call  the cluster, with the
largest summed charge.
The events above (below) the red solid line in Fig.~\ref{fig_neutron} are primarily due to neutrons (gamma rays).  In the figure, the small
dots are data taken with the neutron generator on and the blue crosses are the data taken with a $^{22}$Na source located
between the CsI crystal and neutron detectors. Two 511~keV gamma rays simultaneously emitted from the $^{22}$Na source are
detected by the CsI crystal and one of the neutron detectors.  From these data we estimate that only about 0.3 $\pm$ 0.1(stat.)\% of gamma-induced
events are misidentified as neutron events.

Another criteria for the event selection, called fit quality, required that the signal of the CsI exhibited a decay time that is characteristic of CsI
crystals when fitted with a single exponential function. This criteria removed low energy events triggered by the
long tails of a high-energy signal in the CsI crystal that happened to occur in accidental coincidence with a background signal in the neutron detector.

  The temperature of the experimental environment was controlled to be stable at 18~$^{\circ}$C and an energy calibration
with an $^{241}$Am source was performed every three days during the data taking.

\section{Simulation}
\label{sec:simul}

\begin{figure}[!ht]
\begin{center}
\includegraphics[width=1.0\textwidth,keepaspectratio=true]{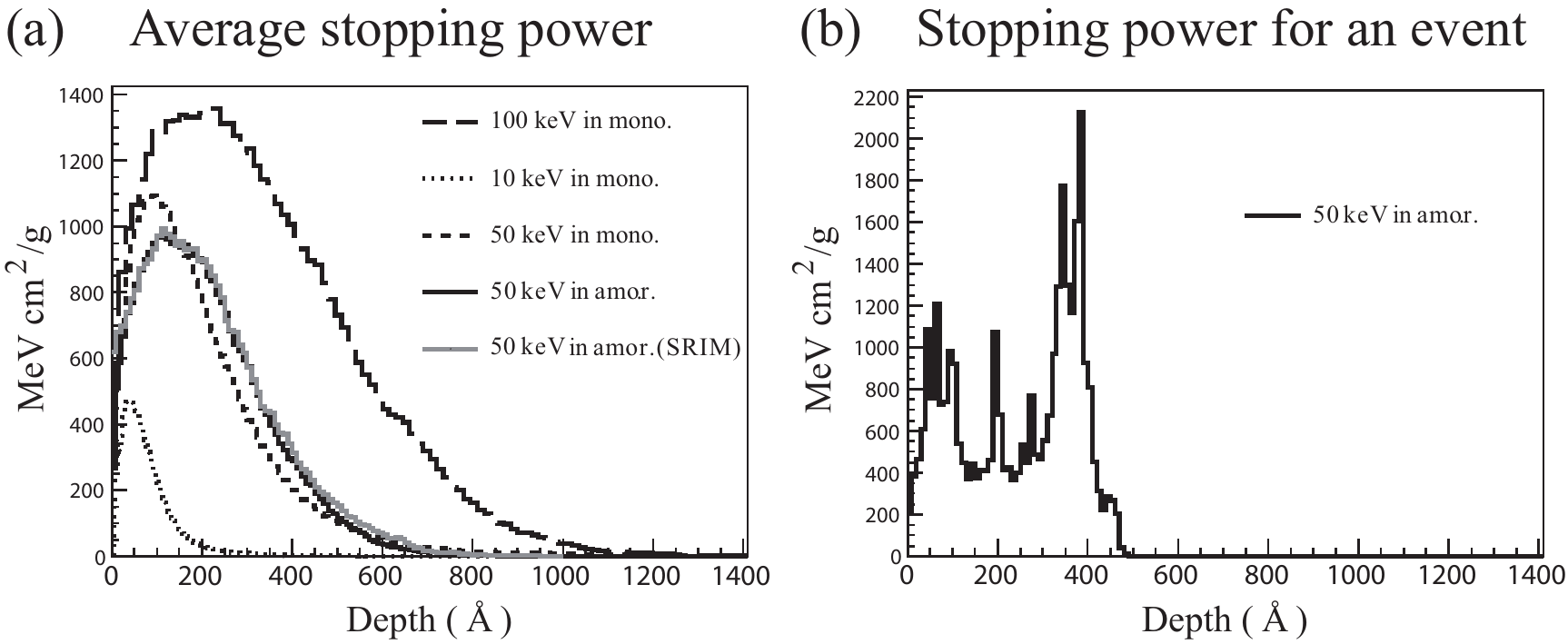}
\caption[] {The electronic stopping power distribution versus the penetration depth. These histograms were obtained from the averaged ionization energy loss at each penetration depth bin of the primary I ion and its recoil ions for all simulated events (a) and for a typical single event (b). The kinetic energy of the primary I ion was initially set as indicated in the figure legends : mono. and amor. indicate that a monocrystalline CsI(Tl) or an amorphous CsI(Tl) crystal was used as the target materials in the simulation, respectively. The gray line histogram in (a) was constructed with a SRIM simulation; the others were obtained with a MARLOWE simulation. }
\label{fig_averageSP}
\end{center}
\end{figure}

A full simulation of the experimental setup was carried out and the results were compared with the experiment data. 
The simulation consists of two parts: the scattering of the neutron beam in the CsI crystal and the subsequent detection in
the neutron detector that is simulated by GEANT4. At each step, the deposited energy is converted to a scintillation light output in keVee units based on a simulation study for recoil ions in a CsI crystal using the MARLOWE simulation code with a modified Birks' formula~\cite{jhlee}. A Monte-Carlo (MC) method is used to construct photoelectron signals for each event that is tuned to reproduce the measured energy spectra with an energy resolution and a detection efficiency that were compatible with experimental observations.

MARLOWE is a binary cascade simulation code for atomic collisions~\cite{marlowe,robinson}. Even though it is not as widely
used as SRIM~\cite{srim} it has the feature of simulating ion transport not only in monocrystalline
or polycrystalline materials, but also in amorphous materials. The MARLOWE program reads an input file that sets parameters that are relevant to the crystal structure, material properties, physics models for nuclear and electronic stopping, selection criteria for target atoms, etc. Using parameter values taken from the papers cited in ref.~\cite{jhlee} with some adjustment, we find that the ions’ range and stopping power obtained from the MARLOWE program are compatible to those determined from SRIM for amorphous CsI(Tl). The only adjustment we made was to the fraction of the nonlocal part in the Oen-Robinson electronic stopping power model~\cite{marlowe}.  In this model, the nonlocal part is independent of the impact parameter between a projectile ion and a target atom, while the local part has dependence on the electron density of the target atom. The nonlocal fraction can be different for various materials. In ref.~\cite{hobler_sim}, the authors used 0.4 to model the ranges of boron ions in silicon.  In this work we use 0.65.

 Figure~\ref{fig_averageSP} shows the electronic stopping power at each penetration depth for Iodine (I) ions obtained from the averaged ionization energy loss in each bin. Each individual event has large fluctuations as shown in Fig.~\ref{fig_averageSP}-(b), but the average of all events is the smooth curve shown in Fig.~\ref{fig_averageSP}-(a). For 50 keV Iodine ions passing through an amorphous crystal, the average electronic stopping power and total penetration range from the MARLOWE simulation are similar to those from SRIM with the parameters mentioned above. However, when the target is changed to a monocrystalline structure with the target atoms moving with thermal vibrations that are centered at each lattice site, the shape of the averaged electronic stopping power distribution becomes more asymmetric and the maximum penetration range increases as shown in Fig.~\ref{fig_averageSP}-(a).

For light recoil ions, such as those commonly considered in scintillation efficiency studies~\cite{murray}, the stopping power is dominated
by the electronic stopping power. However, for heavy
ions, such as Cs or I in a CsI crystal, with recoil energy below some level for each nucleus, the nuclear stopping power due to phonon
creation at each penetration depth bin is larger than the electronic stopping power. Since the scintillation light output is
generated from the electronic stopping power,
we used the electronic stopping power given by the MARLOWE program as an estimate of the scintillation yield.

Once the electronic stopping power is obtained from the MARLOWE
program, we can estimate the differential scintillation light yield per energy deposit from Birks' formula~\cite{birks}: 
\begin{equation}
\frac{dL}{dE} \ = \ \frac{S}{1+kB \frac{dE}{dr}},
\label{eq_birks}
\end{equation}
where $L$ is the generated scintillation yield, $E$ is the energy deposited by the ion in the material, $S$ is the absolute
scintillation yield per unit deposit energy, $kB$ is the Birks' factor, and $\frac{dE}{dr}$ is  the ionization energy
per unit penetration depth.  For stopping powers below 20~MeV$\cdot$cm$^2$/g, we used a modified Birks' formula, 
based on Eq. (11.8a) in ref.~\cite{birks}, which fits the data better~\cite{murray, gwin}:
\begin{equation}
\frac{dL}{dE} \ = \ \frac{k\alpha\frac{dE}{dr}}{1+k\alpha\frac{dE}{dr}} \frac{S}{1+kB\frac{dE}{dr}},
\label{eq_mbirk}
\end{equation}
where the additional term containing $k\alpha$ describes the correlation between the scintillation yield and the number of
electron-hole pairs in the low stopping power region.

As discussed in ref.~\cite{jhlee}, the parameters in the modified Birks' formula were obtained by fitting
Eq.~(\ref{eq_mbirk}) to the measured scintillation efficiency data~\cite{gwin} for the stopping powers of
light charged particles in CsI(Tl) crystals. The parameters determined from this fit are $k\alpha$ = 1~g/MeV$\cdot$cm$^{2}$,
$kB$ = $3.8\times10^{-3}$~g/MeV$\cdot$cm$^{2}$ and $S$ = 1.375~keVee/keV.
Equation~(\ref{eq_mbirk}) was then multiplied by the average ionization energy at each penetration depth and summed for the whole bin to obtain the light output for the CsI crystal as shown in Eq.~(\ref{eq_sum}).
\begin{equation}
L \ =  \ \sum_{i=0}^{max. bin} {\Delta L_i} \ = \ \sum_{i=0}^{max. bin} \frac{k\alpha(\frac{dE}{dr})_i}{1+k\alpha(\frac{dE}{dr})_i} \frac{S}{1+kB(\frac{dE}{dr})_i} {\Delta E_i} ,
\label{eq_sum}
\end{equation}
where $L$ is the total light output, and $\Delta L_i$ and $\Delta E_i$ are the light output and the ionization energy at each penetration depth bin.  

The data used to fix the parameters of the modified Birks' formula were taken with an experimental setup different
from the current one. 
The light output produced by gammas depends upon the DAQ time window because of the presence of slow component of the scintillation, and this leads consequently to different energy resolutions and some non-linearity~\cite{moszynski}.  On the other hand, the scintillation from alphas has smaller slow components than that for gammas and is, therefore, less dependent on the DAQ time window.  As a result, the measured quenching factor for alphas can depend on the gamma reference line that is used in an experiment. The simulated output of Eq.~(\ref{eq_mbirk}) has been adjusted by an additional correction factor of 0.93 in order to reproduce the alpha quenching factors that are consistent with measurement~\cite{jhlee}.  For the current experiment, the correction factor may be different from the one we used, so we introduced another factor, called the shift factor, while retaining the scintillation efficiency function multiplied by 0.93. This shift factor will be applied to the number of photons in the Monte-Carlo method explained later to vary the relative light outputs of nuclear recoils.

\footnotetext{Erratum : DAQ time window of 25~$\mu$s should be changed to 10~$\mu$s on page 3 in ref.~\cite{jhlee}.}

\begin{figure}[!ht]
\begin{center}
\includegraphics[width=1.0\textwidth,keepaspectratio=true]{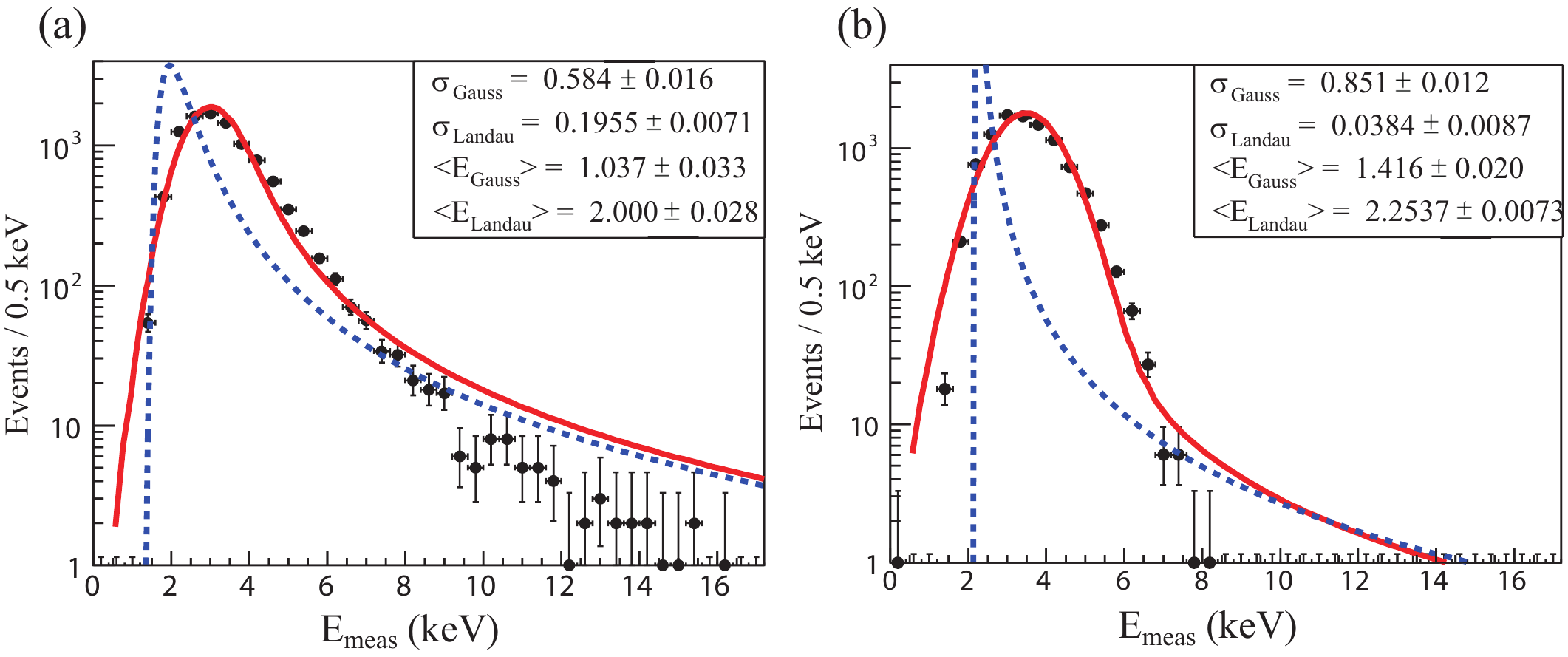}
\caption[] {(color online) Distributions of simulated scintillation light output \emeas{} from the simulation of Cs ions
with 40~keV recoil energy. (a) is for the monocrystalline CsI(Tl) crystal and (b) is for the amorphous CsI. The
solid points are the result of the simulation, the red solid lines are the fitted results for the Landau-Gaussian functions, and the blue dashed lines are the Landau distributions constituting the fitting functions. The
corresponding fit parameters are shown in the inset for each panel. }

\label{fig_marlowe}
\end{center}
\end{figure}

The solid points in Fig.~\ref{fig_marlowe} show the simulated energy distribution (\emeas{}) for Cs ions with a recoil energy of 40~keV in
monocrystalline (a) and amorphous CsI(Tl) material (b) using the MARLOWE program with the modified Birks' formula. Here we varied the initial directions of ions isotropically
within $\pm 20^{\circ}$ with respect to the direction perpendicular to the target plane. The initial positions of the Cs ions were set at lattice sites to simulate the motion of recoil ions inside the crystal.  We fit the simulation results using
Landau-Gaussian functions for the Probability Density Functions (PDFs):
\begin{equation}
\textrm{PDF} = \frac{1}{\sigma _\textrm{\scriptsize L} }{Landau}(E - E_\textrm{\scriptsize L})
\frac{1}{\sqrt{2 \pi} \sigma_\textrm{\scriptsize G}} e^{ (-(E - E_\textrm{\scriptsize G})^{2} / 2\sigma_\textrm{\scriptsize G}^{2} ) },
\label{eq:lg}
\end{equation}
where $E_\textrm{\scriptsize L}$ is the most probable value of the Landau distribution, $E_\textrm{\scriptsize G}$ is the mean value
of the Gaussian distribution for the measured energy, and $\sigma_\textrm{\scriptsize L}$ and $\sigma_\textrm{\scriptsize G}$ are
the standard deviations of the two distributions. This convolution function is calculated by the ROOFIT module in the ROOT
program~\cite{root} and $Landau$ is the Landau distribution function in the ROOT math library. The result of these fits
are shown as solid red curves in Fig.~\ref{fig_marlowe}. 

As is evident in Fig.~\ref{fig_marlowe}, the shape of the PDF for the monocrystalline is different from that for the
amorphous crystal. The enhancement of events in high energy tail region for the monocrystalline case were correlated with the
ions' ranges, so these events are identified as being due to channeling events. These are from recoil ions that are captured in a channel with the help of lattice vibrations and/or the reduction of their energies after several scatterings. The lower the ion energy is, the larger the critical angle or the probability of the channeling. According to Hobler~\cite{hobler}, this continues until the critical approach distance between a projectile ion and a target atom is similar to the channel radius, the half distance between symmetric axes or planes, after which the critical angle goes to zero. Below this energy, ions with any incident angle scatter out of the channel, i.e., dechannel. 
 These correspond to partially channeled events. Full channeling occurs when ions enter a channel at the start of motion and do not escape from the channel until they stop. Since an amorphous
crystal has no symmetry axes or planes, neither partial nor full channeling occurs.

\begin{figure}[!htb]
\begin{center}
\includegraphics[width=0.8\textwidth,keepaspectratio=true]{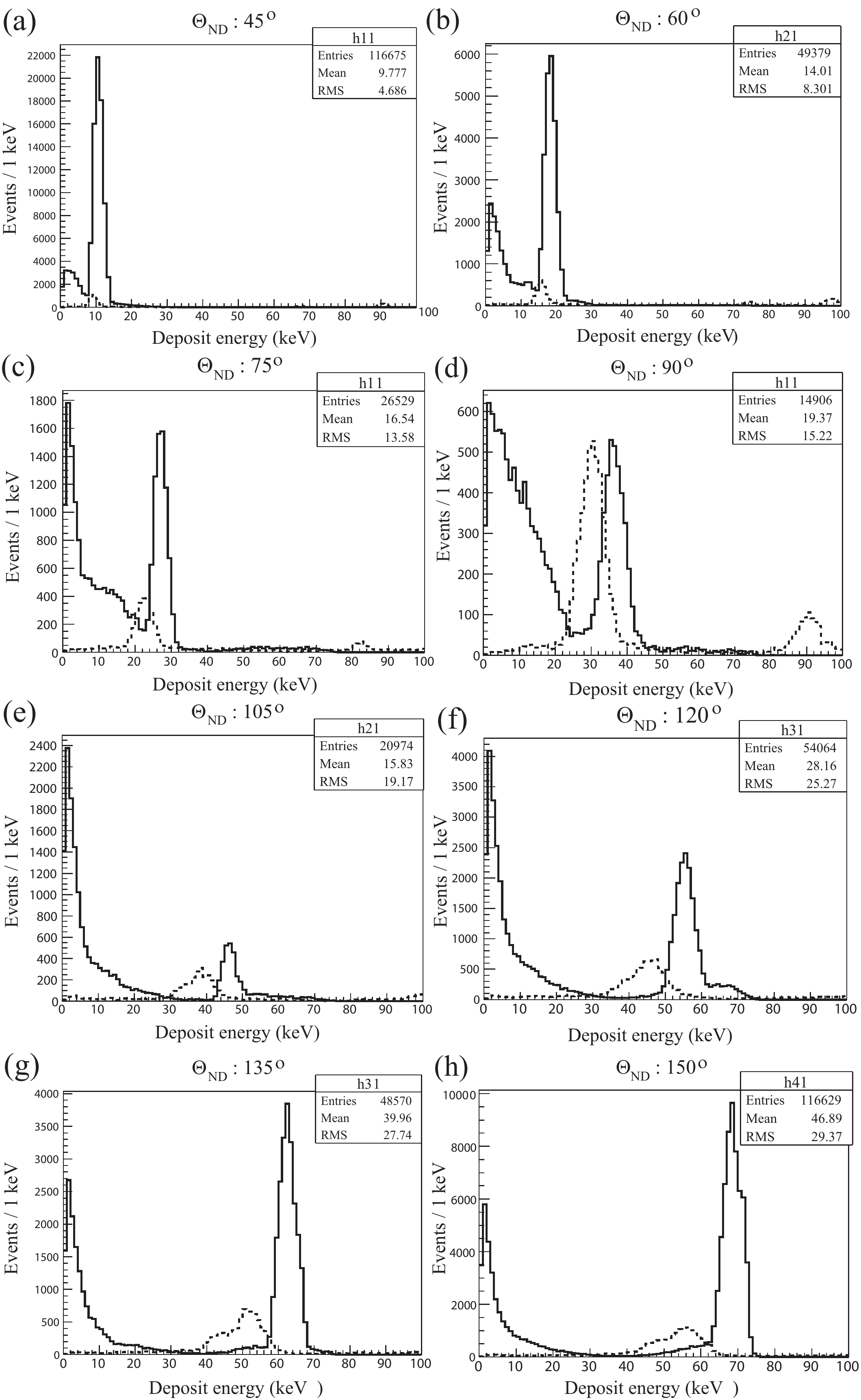}
\caption[] {Recoil energy spectra for single elastic (solid lines) and single inelastic (dashed lines) scattering events of neutrons on Cs or I ions. After fitting each peak with a Gaussian function, we determined the interest region of recoil energy according to the criteria mentioned in the text, and selected events in the measured energy spectra in the simulation to calculate the quenching factor. Only those events have the correct correlation between the angle of the detected neutron and the energy deposited in the CsI crystal by a nuclear recoil. }
\label{fig_er}
\end{center}
\end{figure}

As mentioned above, we used GEANT4 to simulate the crystal response to neutron-recoil events and the quenching
and channeling effect were simulated by the MARLOWE program with the modified Birks' formula. The simulation incorporated the
experimental geometries of Fig.~\ref{fig_setup}.   Since the application of the MARLOWE simulation at each GEANT4 step inside the CsI
crystal is CPU intensive, we accelerated the simulation by producing a template of PDFs as a function of
(\erecoil{},\emeas{}) beforehand. We performed the full MARLOWE simulation for a series of recoil energies relevant to
our experimental setup and determined the parameters of the PDF of Eq.~(\ref{eq:lg}) for each recoil energy. Then, by fitting these parameters
as a function of the recoil energy, a template PDF as a function of \erecoil{} was generated. For each neutron-ion
scattering in the GEANT4 simulation, the PDF is applied for the corresponding recoil energy, and the \emeas{} values is chosen randomly
with probabilities provided by PDF. In the simulation, we used GEANT version 4.9.6.p02 with the NeutronHP model and
G4EmLivermorePhysics builder for the nuclear- and electron-recoil simulations~\cite{geant4}. We excluded events that do not conserve energy-momentum in neutron-target atom inelastic scattering events, and removed some elastic events with some reduction in efficiency. Figure~\ref{fig_er} shows recoil energy spectra deposited in a CsI(Tl) crystal from single scatterings of neutrons on target atoms obtained from GEANT4 simulation, corresponding to neutrons that, after the scattering, enter one of neutron detectors and deposit some energy. Solid lines depict recoil energy spectra from single elastic scatterings and dashed lines are those from single inelastic scatterings. Most of the inelastic scattering events in these spectra correspond to those in which gamma rays escape. However, high energy peaks from additional energy deposited by gamma rays from excited nuclei can be seen in several figures. For the energy spectra from single elastic scatterings, there are unexpected excesses of entries around the expected recoil energy peaks. These events are generated by multiple scatters in the shielding materials after a single scattering in the CsI(Tl) crystal. The relative fraction of multiply scattered events increases as scattering angles where the cross-section of neutron-nucleus single elastic scattering is small. Because of this, we select the single hit events in the GEANT4 simulation for the estimation of quenching factors by limiting the recoil energies to a Region of Interest (ROI) : $ \langle E_{\rm recoil, \rm inelastic} \rangle - 3 \sigma_{\rm recoil, \rm inelastic} < E_{\rm recoil} < \langle E_{\rm recoil, \rm elastic} \rangle + 3 \sigma_{\rm recoil, \rm elastic}$, where $\langle E_{\rm recoil} \rangle$ and $\sigma_{\rm recoil}$ are the Gaussian mean and sigma of single inelastic or elastic scattering peak, respectively.

 After we determined the measured energy, or the electron equivalent energy, from a GEANT4 simulation, we applied a Monte-Carlo method to generate photons corresponding to that energy. In this Monte-Carlo process, we chose a random value that follows a Poisson distribution for a photon yield whose mean number per keV is determined by the gamma calibration.  For energy deposited by gamma rays, the resolution was equivalent to that of gamma calibration data using an $^{241}$Am source, while for recoil ions, the resolution was a combined result of a random choice from a Poisson distribution and the photo-electron charge distribution in the experiment.  After the Monte-Carlo data are generated, we apply the two-fold trigger condition requirements as a hardware cut and the fit quality requirements as a software cut to mimic those that were applied to the experimental data. This Monte-Carlo process is repeated with the application of a shift factor to the photon yield from a recoil ion determined from the modified Birks' formula and the gamma calibration. Since the fitted formula is based on the other experimental data, the repetition is a way to find a function that describes our data better. In the next section, we present the best fit $E_{\rm meas}$ spectra for data and simulation at each setup of the scattering angle using the most probable shift factor obtained from the minimum negative log likelihood fit method.

\begin{figure}[!htb]
\begin{center}
\includegraphics[width=0.9\textwidth,keepaspectratio=true]{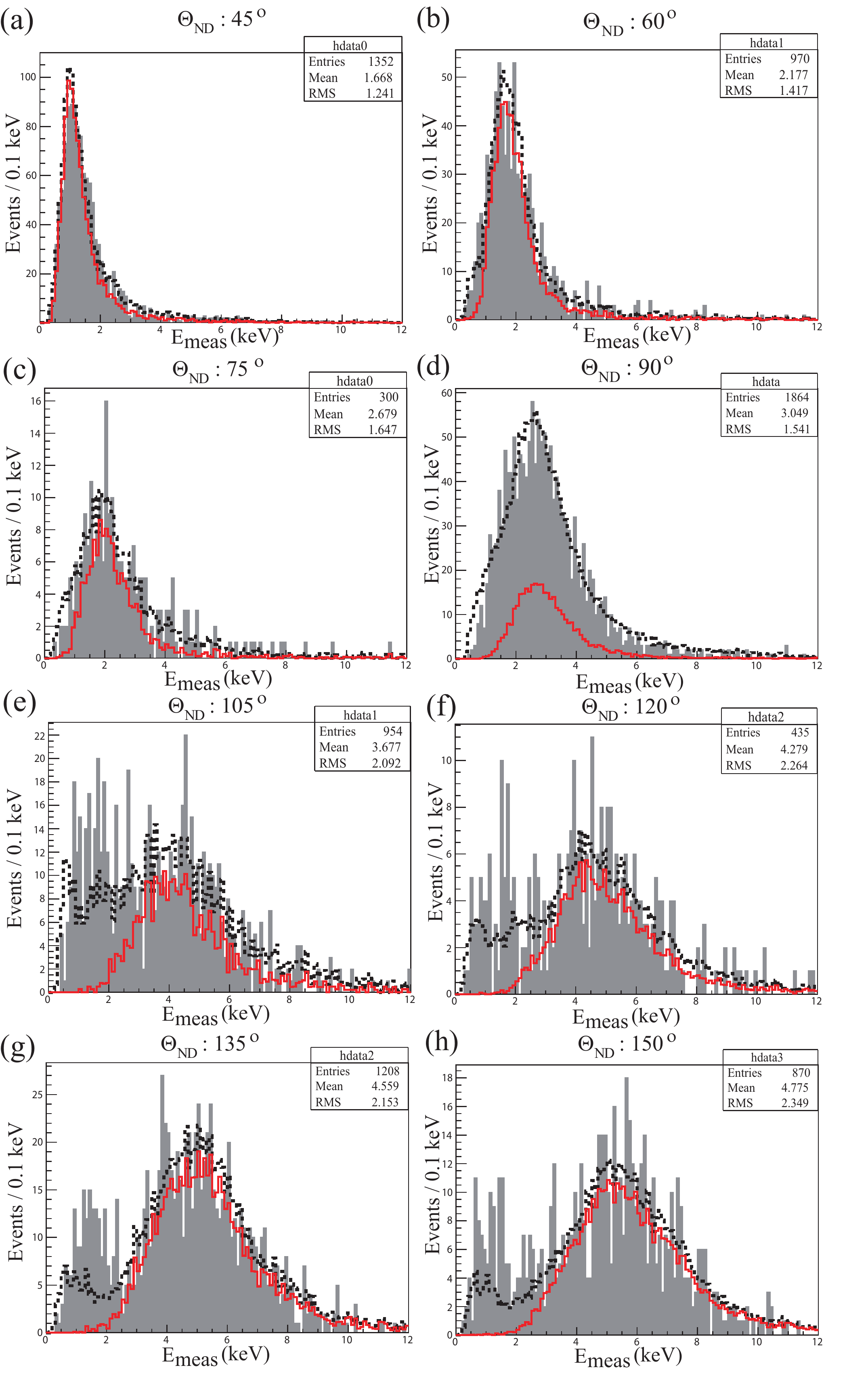}
\caption[] {(color online) \emeas{} distributions for nuclear recoil events tagged by neutron detectors of different neutron scattering
angles of Setup I (shaded histogram) and the simulated distributions from the GEANT4 simulation (dashed lines) below 12~keV.
The red histograms show \emeas{} for single MC events that are within the recoil energy regions of interest (ROI).
}
\label{fig_meas}
\end{center}
\end{figure}

\begin{figure}[!htb]
\begin{center}
\includegraphics[width=1.0\textwidth,keepaspectratio=true]{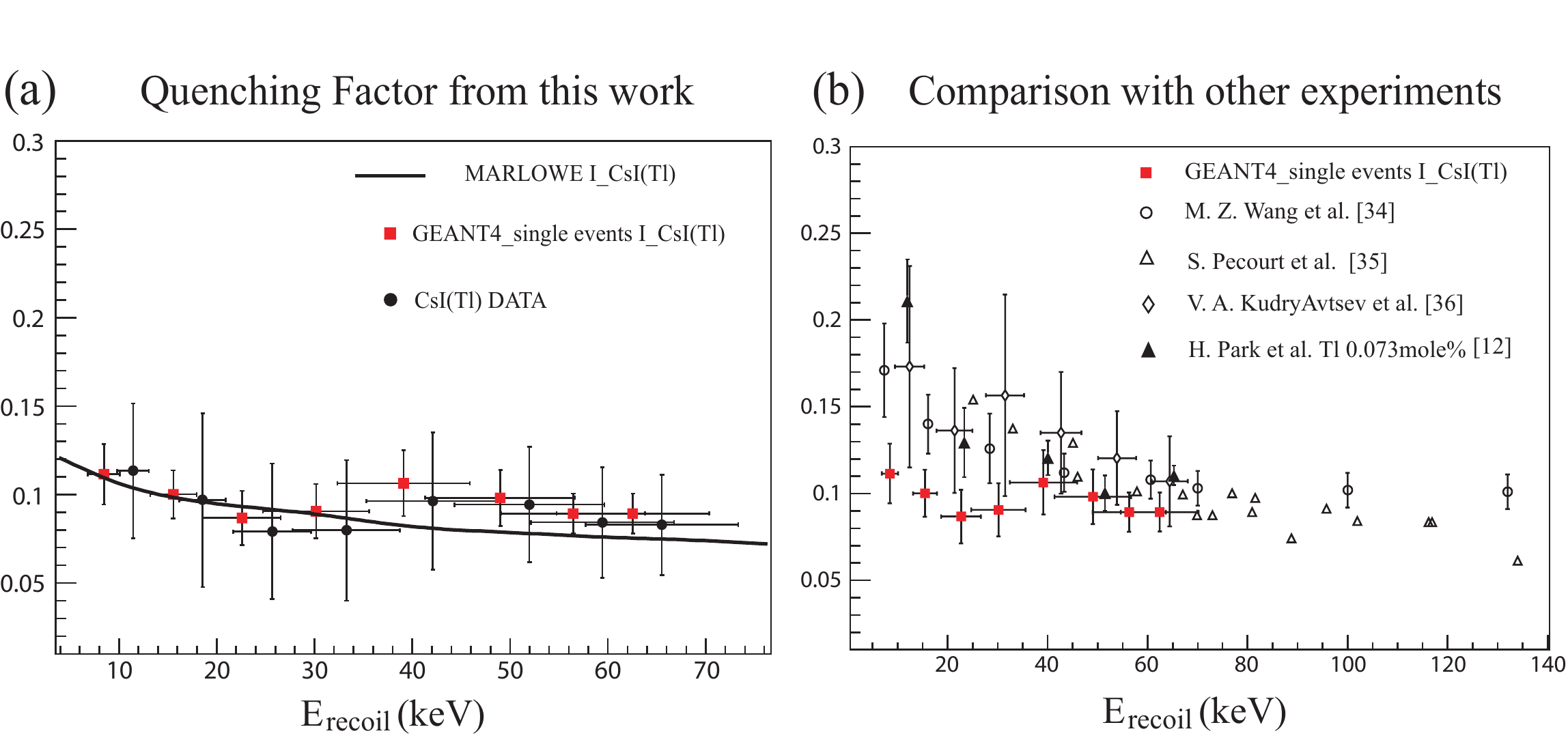}
\caption[] {(color online) The CsI(Tl) quenching factors as a function of recoil energy. (a) shows the quenching factors of this work, and (b) shows the comparison of quenching factors of this work and previous experiments. The black circles and red squares in (a) are obtained from fitting Gaussian functions to the experimental $E_{\rm meas}$ distributions and those from red solid lines in Fig.~\ref{fig_meas}. The red squares are shifted by -2 keV along the x-axis for clarity. The line in (a) is a spline function obtained from simulation quenching factor from MARLOWE simulation depicted in Fig.~\ref{fig_marlowe}.}
\label{fig_qf}
\end{center}
\end{figure}

\begin{figure}[!hbt]
\begin{center}
\includegraphics[width=1.0\textwidth,keepaspectratio=true]{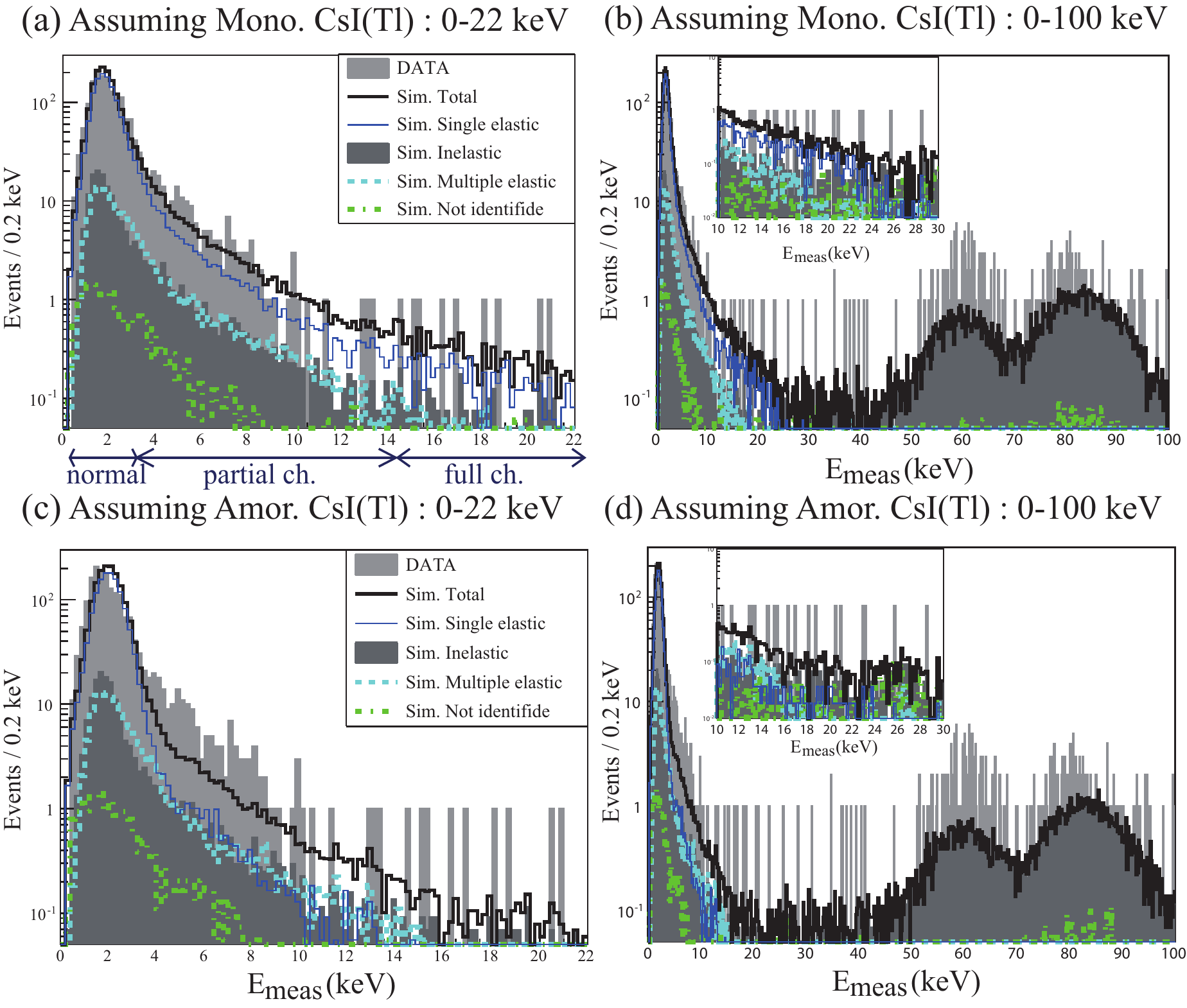}
\caption[] {(color online) Comparison between the experimental and simulated \emeas{} distributions for nuclear recoil events
tagged by neutron detectors at $\Theta_\textrm{\scriptsize ND}$ = 60$^{\circ}$ (shaded histograms) for Setup II.
The simulated distributions (black open histograms) are decomposed into single elastic (narrow blue line), inelastic (dark-gray shaded), multiple scatterings
(light-blue dashed line) and unidentified events (green dash-dotted line). (a) is for the monocrystalline CsI(Tl), (c) is for the amorphous CsI(Tl), and arrows below the x-axis in (a) indicate approximate energy regions categorizing events according to Table~\ref{table1}. (b) and (d) are zoomed
views of the $x$-axes of (a) and (c) and insets in (b) and (d) show events around the full channeling region. }
\label{fig_gmeas}
\end{center}
\end{figure}

In summary, we have simulated the distributions of measured CsI(Tl) energy for several recoil energies using a MARLOWE-based simulation together with a modified Birks' formula Eq.~(\ref{eq_mbirk})~\cite{birks, jhlee}.
The resultant distributions are fitted with Landau-Gaussian functions that are incorporated into the GEANT4 simulation. And then, by a Monte-Carlo method the signal shape per each event was reproduced, and the simulated $E_{\rm meas}$ spectra were completely reproduced after applying the selection-efficiency corrections.   

This method to obtain \emeas{} for each \erecoil{} can be compared to the work of Bernabei~et~al.~\cite{bernabei_channeling},
where \emeas{} from channelling events was obtained using a simple model without full simulation.
Previous work by Hitachi~\cite{hitachi} and Tretyak~\cite{tretyak} simply used the stopping power and the scintillation efficiency
function to obtain the mean of \emeas{} or, equivalently, the quenching factors.  The work reported here differs from
the previous work in that the simulation of neutron-ion scatterings and ion-ion scatterings are fully incorporated using
the GEANT4 and MARLOWE programs, so the simulated distributions can be directly compared with experimental
measurements of both elastic and inelastic scattering, including channeling effects.

\section{Result and discussion}
\label{sec:result}

Figure~\ref{fig_meas} shows the experimentally measured (shaded histogram) and the simulated (dashed line) $E_{\rm meas}$
distributions for nuclear recoil events tagged by each neutron detector in experiment with Setup I.

Low energy peaks near 1~keV in the simulated spectra in Fig.~\ref{fig_meas}-(e)-(h) are found to come from neutrons that scatter in the lead
blocks around the CsI detector after scattering in the crystal. For Fig.~\ref{fig_meas}-(a)-(d), those events contaminate the single hit events with recoil energy in the ROI, which is reflected in discrepancies between the red solid curves denoting the simulated spectra of single hit events within ROI and the dashed lines that indicate the distribution of all of the simulated events. Furthermore, bumps in right hand side of the peaks from single elastic scatterings shown in Fig.~\ref{fig_meas}-(e) and (f), the multiply scattered events with shielding material, distort the measured energy distributions, since their $E_\textrm{\scriptsize{meas}}$ are within the sigma values of the Gaussian distributions for the single elastic events.  This implies that the quenching factors estimated from the $E_\textrm{\scriptsize{meas}}$ spectra from the experiment may contain systematic errors. So in order to estimate the correct quenching factor from the simulation, we consider only single hit MC events within the ROI, which are shown as solid red histograms in Fig.~\ref{fig_meas}. Since we applied a stronger condition to select the single hit events for the setup of the $\Theta_\textrm{\scriptsize ND}$ = 90$^\circ$ due to larger solid angle of incident neutrons than in other experimental setups, the fraction of single hit events within the ROI is smaller than for other cases in Fig.~\ref{fig_meas}-(d).

 As mentioned in the previous section, the most probable shift factor for the photon yield was applied in order to correctly reproduce the $E_{\rm meas}$ fitted value from the data. For eight different $\Theta_\textrm{\scriptsize ND}$ values, the mean and sigma of the most probable shift factor was 1.133 and 0.105. For $\Theta_\textrm{\scriptsize ND}$ = 75$^\circ$ and 105$^\circ$ in Fig.~\ref{fig_meas}-(c) and (e), the shift factors are larger that the rms spread ($\sigma$). 
These fluctuations may be due to the contamination of multiply scattered neutrons in the lead blocks around neutron detectors that are not included in our GEANT4 simulation, and the relatively low cross-sections of the neutron-nucleus elastic scattering at those angles. Actually we expect the shift factor might be smaller than the one we used because of non-linearity between the 59.54 keV reference line we used, and 662 keV gammas used in ref.~\cite{gwin}. However, there may be additional factors, for example, differences in the relative scintillation efficiencies for alpha and nuclei~\cite{murray}; these will be studied later.

  The black circles in Fig.~\ref{fig_qf}-(a) depict the measured raw quenching factors from this experiment. For the calculation of the quenching factors, we set the denominator in Eq.~(\ref{eq:QF}) to the average $E_{\rm recoil}$ within the ROI discussed in section~\ref{sec:simul} with the error taken to be its standard deviation. The numerator is the Gaussian mean value of $E_{\rm meas}$ in the data, where the fit range was determined by the category of the normal scattering represented in Table~\ref{table1}. 
 The quenching factor from MARLOWE was represented as a spline function using the Gaussian fit mean for the $E_{\rm meas}$ spectrum shown in Fig.~\ref{fig_marlowe}-(a) and given by the MARLOWE program with the modified Birks' formula for mono-energetic ions. This shows energy dependent quenching factor without considering the shift factor in our scintillation model. 
 With the red histograms in Fig.~\ref{fig_meas} from the selected events within ROI in the reproduced $E_{meas}$ spectra after applying the shift factor, we obtained the corrected quenching factors as red square points in Fig.~\ref{fig_qf}. The lower value at $\Theta_\textrm{\scriptsize ND}$ = 75$^\circ$ and higher values at $\Theta_\textrm{\scriptsize ND}$ = 105$^\circ$ are conjectured to be due to a larger contamination of multiply scattered neutron events in the lead blocks around the neutron detectors that is not included in our GEANT4 simulation, and associated with an initial scattering angle in the CsI crystal that is different from the nominal value.

Figure \ref{fig_gmeas} shows the distribution of measured energies, \emeas{}, obtained in Setup II with the same CsI
crystal.  Here the contributions from the simulated \emeas{} due to single elastic scatters, multiple scatters,
and inelastic scatters in the CsI crystal are indicated separately. The single elastic scattering events are dominant in the low energy
peak region above 1~keV. From the figure, it is clear that the simulation result with the amorphous CsI crystal cannot
explain the experimental data (Fig.~\ref{fig_gmeas}-(c)), while the simulation with the monocrystalline CsI 
reproduces the experimental data well for all energy region (Fig.~\ref{fig_gmeas}-(a)) as expected in Fig.~\ref{fig_marlowe}.

Figures~\ref{fig_gmeas}-(b),(d) show the measured energy distributions over an extended energy range, where two
inelastic scattering peaks; 57.6~keV gamma rays from the deexcitation of the first excited state of I and 79.6 and
81~keV gammas from the deexcitation of the first excited state of Cs, with added contributions of the recoil energy,
can be seen.  These experimental distributions are compared with the simulation after normalizing by the total number of events
in a limited energy region of normal scattering as indicated in Table~\ref{table1} and used in Fig.~\ref{fig_meas}. There is an excess of events in Fig.~\ref{fig_gmeas}-(b),(d) above 30 keV in x-axis
 that we attribute to gamma-induced events in the neutron detector that are
misidentified as neutron events by the PSD; when we consider the neutron PSD data shown in Fig.~\ref{fig_neutron}, the total number of
gamma ray events in data was about 15,000 per each neutron detector and with consideration of the 0.3 $\pm$ 0.1(stat.)\% PSD misidentification fraction
determined from the $^{22}$Na calibration run, the origin of excess of events in \emeas{} can be accounted for. The insets in Fig.~\ref{fig_gmeas}-(b),(d) show events in the range from 10 keV to 30 keV around the full channeling region, where the simulation with an assumption that the detector crystal is amorphous hardly reproduce data due to the deficit of channeling events in the simulation.

\begin{table}[!t]
\renewcommand{\arraystretch}{1.3}
\caption{Event categories depending on the $E\rm_{meas}$}
\label{table1}
\centering
\begin{tabular}{|c|c|}
\hline
Category & Energy range\\
\hline
\hline
Normal scattering & $ E_\textrm{\scriptsize meas} < ~E_\textrm{\scriptsize peak} + 3 \sigma_\textrm{\scriptsize peak}$\\
\hline
Partial channeling & $E_\textrm{\scriptsize peak} + 3 \sigma_\textrm{\scriptsize peak} ~\leq E_\textrm{\scriptsize meas} \leq    \langle E_\textrm{\scriptsize recoil} \rangle - \, 3\sigma_\textrm{\scriptsize recoil}$\\
\hline
Full channeling & $ \langle E_\textrm{\scriptsize recoil} \rangle - \, 3 \sigma_\textrm{\scriptsize recoil} < E_\textrm{\scriptsize meas} <  \langle E_\textrm{\scriptsize recoil} \rangle \, + 3 \sigma_\textrm{\scriptsize recoil} $\\
\hline
\end{tabular}
\end{table}
 Table~\ref{table1} shows the event categories that we used for the estimation of quenching factors and channeling effects and they are indicated by arrows below the x-axis in Fig.~\ref{fig_gmeas}-(a). $\sigma_\textrm{\scriptsize peak}$ and $\sigma_\textrm{\scriptsize recoil}$ are the standard deviations of the
peak of normal nuclear recoils in the experimental data and the peak of the \erecoil{} in the GEANT4 simulation,
respectively. Since the nuclear recoil peak cannot be fitted successfully with a simple Gaussian ansatz due to the
asymmetric shape of the data, we set the mean value of \emeas{} to be $E_\textrm{\scriptsize peak}$.  We estimated the partial
and full channeling fractions with statistical errors in the data shown in Fig.~\ref{fig_gmeas} as 17.04 $\pm$ 1.07(stat.)\% and 0.75 $\pm$ 0.20(stat.)\% of the total events for
all three categories. However, there are contaminations from gamma-induced events,
multiply scattered events and inelastic scattering events as the simulated spectra shown in Fig.~\ref{fig_gmeas}. The partial and full channeling fractions estimated from the simulation are 5.26\% and 0.55\% after removing multiple and inelastic scattering events in the CsI crystal and single hit events with multiple scatters in the shielding material that are estimated from the energy spectrum for the amorphous CsI crystal simulation shown in Fig.~\ref{fig_gmeas}-(c). In any case, these numbers are too small to affect the quenching factor.

 In the simulation, we used two tunable parameters.  One is the nonlocal content for matching electronic stopping power distributions for each penetration depth for SRIM and MARLOWE. And the other is the correction factor for matching measured energy spectra of measurement and simulation, which is resultantly 5\% increase in the scintillation efficiency function from the calculation that 0.93  is multiplied by the mean shift factor of 1.133. Although there are uncertainties in this model, it is a good way to estimate the channeling effect on the quenching factor by means of simulation for the motions of recoil ions in a crystal. 

\section{Conclusion}
\label{sec:conclusion}
We measured the quenching factors of a CsI(Tl) crystal to study channeling effects and to analyse the recoil energy
spectra expected from WIMP interactions. By comparing experimentally measured \emeas{} distributions with simulations,
we find that the simulation method of \emeas{} using a scintillation efficiency function of the electronic
stopping power in the CsI crystal reproduces the data. The quenching factors reported here are lower than the previous
measurements. The discrepancies could be due to the use of different gamma sources for the
energy calibration and the trigger inefficiencies for nuclear recoils at low energy~\cite{collar} due to smaller photon yield in the previous measurements.
More studies on the quenching factors are in progress that address these questions. We were able to estimate the
fraction of partial channeling as 5.26\% with the simulation, contributing to a tail of events with
enhanced \emeas{}, while the 0.55\% fraction of fully channeled events is at a nearly insignificant level. While this value may have a large uncertainty, however consistent with the result of Bozorgnia et al. that the maximum fraction for the full channeling is about 2\% at a recoil energy of 20 keV at room temperature.

\section{Acknowledgement}
This work was supported by WCU program (R32-10155) and Basic Science Research Grant (KRF-2007-313-C00155) through the
National Research Foundation funded by the Ministry of Education, Science and Technology of Korea. J. H. Lee would
like to thank Stefano Scopel for an insightful discussion.

\end{document}